\documentclass[11pt,a4paper]{article}
\usepackage[utf8]{inputenc}
\usepackage[round]{natbib} 
\usepackage{epic,eepic,epsfig,amsmath,amssymb,amsthm,enumitem,slashbox}
\usepackage{t1enc,tabularx,subcaption,graphicx}
\usepackage{array}
\usepackage{multirow}
\usepackage{blindtext}
\usepackage[francais,english]{babel}

\usepackage{caption}

\usepackage{color}

\usepackage{bbm}

\usepackage{hyperref}
\def\virgp{\raise 2pt\hbox{,}}

\renewcommand{\geq}{\geqslant}
\renewcommand{\leq}{\leqslant}
\def\N{{\mathbb N}}

\def\R{{\mathbb R}}

\def\virgp{\raise 2pt\hbox{,}}
\def\cdotpv{\raise 2pt\hbox{;}}

\def\1{\mathbbm{1}}

\theoremstyle{remark}
\newtheorem{remark}{Remark}[section]

\theoremstyle{definition}

\theoremstyle{definition}

\theoremstyle{definition}

\newcolumntype{M}[1]{>{\centering}m{#1}}

\begin{document}

\title{Keynesian Beauty Contest in Morocco's Public Procurement Reform}

\vskip 0.5cm

\author{Nizar Riane$^\dag$}

\maketitle

{\centerline{Faculty of Law, Economic and Social Sciences - Agdal,}
\centerline{$^\dag$ Mohammed V University in Rabat,} 
\centerline{Avenue des Nations-Unies, B.P. 721 Rabat, Morocco}\centerline{Nizar.Riane@gmail.com}}

\vskip 1cm

\begin{abstract}
This paper examines the recent reform of Morocco's public procurement market through the lens of Keynesian beauty contest theory. The reform introduces a mechanism akin to a guessing-the-average game, where bidders attempt to estimate a reference price, which in turn impacts bidding strategies. We utilize this setup to empirically test key hypotheses within auction theory, specifically the roles of common knowledge and bounded rationality. Our findings indicate potential manipulation risks under the current rules, suggesting that a shift to a median criterion could improve robustness and reduce the likelihood of manipulation. This work contributes to the broader understanding of strategic interactions in procurement and offers a foundation for future research on improving fairness and efficiency in public contract allocation.
\end{abstract}

\vskip 1cm


\vskip 1cm


\vskip 0.5cm

\section{Introduction}

\hskip 0.5cm Competing for service rights through bidding represents a relatively pure form of competition. In a first-price sealed-bid auction within the government procurement market, each bidder submits a sealed offer unknown to the others. The lowest bidder wins the right to provide the service at the price they submitted.\\

L. Friedman's analysis of optimal bids in competitive bidding scenarios stands as one of the seminal works on first-price sealed-bid auctions \citep{Friedman1956}. Friedman employs a probability of winning, which is estimated using previous "bidding patterns" of potential competing bidders. For contract bidding, the estimated probability distribution of the cost to fulfill the contract is utilized.\\

Building on Friedman's work, P. Milgrom developed the concept further through the lens of Bayesian Nash equilibrium \citep{Milgrom1982}. He demonstrated that a player's optimal bid corresponds to the expected value of the highest bid from other participants, conditional upon those bids being lower than his own.\\

Vickrey's contributions have also profoundly influenced auction theory and found applications in diverse fields \citep{Vickrey1961}. The Vickrey auction is widely regarded as a pivotal benchmark in auction design, serving as a standard for evaluating the efficiency of alternative auction formats.\\

This study focuses on the Moroccan regulations governing public procurement. Until 2023, the Moroccan public procurement market adhered to the first-price sealed-bid auction model \citep{DECRET212349}, where the attribution of public works and services contracts was based on selecting the lowest sealed bid.\\

In March 2023, a reform of public procurement was introduced \citep{DECRET222431}, the announced objectives are to preserve the principles of:
\begin{enumerate}
\item freedom of access to public procurement markets;

\item equal treatment of competitors;

\item guarantee of the rights of competitors;

\item Transparency in the choices of the contracting authority.\\
\end{enumerate}

The reform introduces a novel approach to awarding contracts, shifting to a system where the winning offer is determined through an average-guessing contest. According to decree 2-22-431 \citep{DECRET222431}, the best offer is the closest to a reference price $P$ by default, after eliminating abnormally high and low bids. This reference price is calculated using a weighted average that reflects the cost estimates provided by the contracting authority.\\

This rule encourages behaviors compatible with the Keynesian theory of beauty contests. As expressed by John M. Keynes in his General Theory of Employment, Interest and Money \citep{Keynes1936}: "It is not a case of choosing those [faces] that, to the best of one's judgment, are really the prettiest, nor even those that average opinion genuinely thinks the prettiest. We have reached the third degree where we devote our intelligences to anticipating what average opinion expects the average opinion to be. And there are some, I believe, who practice the fourth, fifth and higher degrees."\\

The Keynesian beauty contest was formalized by Herv\'{e} Moulin in 1979 \citep{Moulin1982}. A similar problem was represented by the Guess $p$ of the average game introduced by A. Ledoux \citep{Ledoux1981} and experimented by R. Nagel \citep{Nagel1995} in the context of the cognitive hierarchy theory. R. Nagel's experiments showed that many people make mistakes and do not assume common knowledge of rationality. She proposed an alternative theory of boundedly rational behavior in which the "depths of reasoning" are important. The results indicate that reasoning order $1$ and $2$ play a significant role.\\

Recent research underscores the role of bounded rationality and strategic interactions. Mauersberger, Nagel, and B\"{u}hren \citep{Mauersberger2020} emphasize that bounded rationality, within the Keynesian beauty contest framework, effectively bridges theoretical models with real-world economic applications. Their work investigates the integration of behavioral elements into macroeconomic models to enhance our understanding of economic behaviors.\\ 

Further expanding on bounded rationality, Lin and Palfrey \citep{Lin2024} introduced the dynamic cognitive hierarchy (DCH) model, capturing evolving player beliefs in sequential interactions. Their findings illustrate how belief adjustments influence strategic outcomes in extensive-form games, emphasizing the role of cognitive sophistication in real-world applications.\\

Addressing the challenge of information asymmetry in procurement, Bodendorf, Hollweck, and Franke \citep{Bodendorf2022} utilized a Stackelberg game model to show how supplier information asymmetry can impede effective purchasing decisions. Their model aids in mitigating these discrepancies, facilitating better negotiation strategies and price determinations in digital procurement contexts.\\

Privacy concerns also influence strategic choices in Keynesian contests. Elzayn and Schutzman \citep{Elzayn2019} introduced the "price of privacy," examining how agents' attempts to conceal personal information can reduce social welfare by impairing coordination. Their study quantifies the trade-offs between privacy and efficiency, underscoring the costs of information asymmetry in collective decision-making.\\

A recent study by Cisternas and Kolb \citep{Cisternas2024} explores dynamic signaling in scenarios where senders are unaware of the signals generated by their actions. Their findings on linear-quadratic-Gaussian games, which address higher-order uncertainty, emphasize the importance of second-order beliefs for information transmission, with implications for macroeconomics, reputation, and trading.\\

Cognitive reflection also plays a crucial role in strategic alignment within games, as shown by Ballester, Rodriguez-Moral, and Vorsatz \citep{Ballester2024}. Their extended level-k model reveals that participants with higher cognitive reflection scores tend to perform closer to Nash equilibrium, highlighting cognitive reflection's impact on strategic decision-making.\\

Lastly, Anufriev, Duffy, and Panchenko \citep{Anufriev2024} introduce the Individual Evolutionary Learning (IEL) model, which simulates human behavior in repeated beauty contest games, closely replicating experimental data and providing insights into learning dynamics in competitive environments.\\

In public procurement contexts, Shanglyu Deng examines "Speculation in Procurement Auctions," revealing how speculation affects auction outcomes, suggesting that while it may harm overall auction efficiency, it can benefit sellers at the auctioneer's expense \citep{Deng2023}.\\

In a 2020 study published in the Journal of Regulatory Economics, Bedri Kamil Onur Tas investigates the relationship between public procurement regulation quality and competition, finding that higher regulatory quality enhances bidder competition and leads to more cost-effective contracts \citep{Tas2020}.\\

This paper contributes to the empirical study of cognitive hierarchy in Keynesian beauty contests, focusing on interactions between companies and investors in the Moroccan public procurement market.\\

In section \ref{The Keynesian Beauty Contest in Moroccan Public Procurement}, we analyze the Moroccan public procurement game under the new reform. Section \ref{Cognitive hierarchy theory and limited rationality} empirically tests the hypothesis of common knowledge using data from public procurement tender results. Finally, Section \ref{Moroccan public markets principles and manipulations} explores coalition formation possibilities under the new rules and proposes a solution based on the median.

\vskip 0.5cm

\section{The Keynesian Beauty Contest in Moroccan Public Procurement\label{The Keynesian Beauty Contest in Moroccan Public Procurement}}

\hskip 0.5cm In this article, we focus on a sealed-bid auction. Let $N \geq 2$ denote the number of bidders. Each bidder $i \in \{1, \ldots, N\}$ submits a sealed bid without knowledge of the competitors' offers. To ensure participation, the bid value $x_i$ of each bidder $i$ must be constrained within specific bounds: $A \leq x_i \leq B$, where $A$ and $B$ are predefined constants in $\R$.\\

In the new reform of the Moroccan procurement market \citep{DECRET222431}:
\begin{enumerate}
\item An offer is deemed excessive if it exceeds the contracting authority's price estimate by more than twenty percent (20\%) for contracts related to works, supplies, and services, excluding studies.
\item An offer is classified as abnormally low if it is more than twenty percent (20\%) below the contracting authority's price estimate for works contracts or more than twenty-five percent (25\%) below the estimate for supply and service contracts, excluding studies.
\end{enumerate}

\vskip 0.5cm

\subsection{The Reference Price Guessing Game\label{The Reference Price Guessing Game}}

\hskip 0.5cm We first consider a simplified version of the Moroccan public procurement reform, referred to as the "Reference Price Guessing Game." In this game, the objective is for players to guess the reference price $P$, defined by the formula
\begin{align}
\label{eqn:1}
P &= \frac{E + \displaystyle \frac{\sum_{i=1}^N x_i}{N}}{2}
\end{align}

\noindent where $E$ is the cost estimate of the public procurement, established by the contracting authority.\\

This game is a variant of the Keynesian beauty contest. It can be shown that the Nash equilibrium is given by $E$ through the iterated elimination of weakly dominated strategies, assuming common knowledge\footnote{An event is considered "common knowledge" if, in a given state, each individual knows the event, each individual knows that all other individuals know it, and this knowledge continues recursively: each individual knows that all other individuals know that all individuals know it, and so on \citep{Osborne1994}.}.\\

Define the map $\text{Fix} \, : \R \to \R $ such that $\text{Fix}(Z)$ is the solution to the equation
\begin{align}
\label{eqn:2}
x&=\frac{E+\displaystyle\frac{(N-1)Z+x}{N}}{2}
\end{align}

This can be expressed as
\begin{align}
\label{eqn:3}
\text{Fix}(Z)&=\frac{N\,E+(N-1)Z}{2N-1}
\end{align}

Define the recurrent sequence $\text{Fix}^{n+1}(Z)=\text{Fix}(\text{Fix}^n(Z))$, with $\text{Fix}^{0}(Z)=Z$ for $Z\in\R$. One can express recursively
\begin{align}
\label{eqn:3bis}
\text{Fix}^{n}(Z)&=\frac{\left((2N-1)^n-(N-1)^n\right)\,E+(N-1)^n Z}{(2N-1)^n}
\end{align}

This sequence is a contraction that converges to its unique fixed point $E$. By weak domination argument:

\begin{enumerate}
\item Step 1: If all the players choose the upper bound $B$, then any player's offer $x_i$ will exceed the reference price if $x_i>\text{Fix}(B)$. Conversely, if all players choose the lower bound $A$, any player's offer $x_i$ will fall below the reference price if $x_i<\text{Fix}(A)$. Thus, any offer outside the interval \mbox{$\displaystyle\left[\text{Fix}(A),\text{Fix}(B)\right]$} is weakly dominated, establishing $\text{Fix}(A)$ and $\text{Fix}(B)$ as new psychological barriers.\\

\item Step 2: Given the conclusions from Step 1 and the new rational interval \mbox{$\displaystyle\left[\text{Fix}(A),\text{Fix}(B)\right]$}, a similar argument shows that the new rational interval becomes \mbox{$\displaystyle\left[\text{Fix}^2(A),\text{Fix}^2(B)\right]$}.\\

\item $\hdots$\\

\item Step $n$: the rational interval is \mbox{$\displaystyle\left[\text{Fix}^n(A),\text{Fix}^n(B)\right]$}.\\

\item $\hdots$\\

\item At the limit: The rational interval is reduced to \mbox{$\{E\}$}, which is the Nash equilibrium.
\end{enumerate}

The argument above assumes that each player knows the number of participants $N$, which is an unrealistic hypothesis. To address this issue, we set $A_{0}=A$ and $B_0=B$, and define the recurrent sequences $A_n=\displaystyle\frac{(2^n-1)E+A}{2^{n}}$ and $B_n=\displaystyle\frac{(2^n-1)E+B}{2^{n}}$. This formulation allows us to apply an alternative iterated elimination of weakly dominated strategies (see, for example, \citep{Kocher2014}):

\begin{enumerate}
\item Step 1: If all the players choose simultaneously the upper bound $B$, or the lower bound $A$, the reference price will remain within the interval \mbox{$\left[A_1,B_1\right]$}. Consequently, any financial offer outside this range is weakly dominated, with $A_1$ and $B_1$ establishing new psychological barriers.\\

\item Step 2: By a similar argument, any offer outside the rational interval \mbox{$\left[A_2,B_2\right]$} is also weakly dominated.\\

\item $\hdots$\\

\item Step $n$: Any strategy outside the rational interval \mbox{$[A_n,B_n]$} is weakly dominated.\\

\item $\hdots$\\

\item At the limit, the rational interval converges to the singleton $\{E\}$, forming the unique Nash equilibrium.
\end{enumerate} 

At $E$, no player has an incentive to change their strategy. Suppose there exists a Nash equilibrium $X=(x_1,\hdots,x_N)$ that differs from $E$. If at least one player is not a winner in this equilibrium, they would have an incentive to adjust their bid, thereby altering the reference price $P$. The only scenario where this does not occur is when $x_1=x_2=\hdots=x_N=x$, which leads to the equation $\displaystyle x=\frac{E+x}{2}$, This equation has a unique solution $x=E$.

\vskip 0.5cm

\subsection{The best response dynamic of the reference price guessing game}

\hskip 0.5cm What occurs when the game is repeated? Consider players with limited memory, who base their strategies on the outcome of the previous play. At each step $n$, , the optimal strategy $x_{in}$ for player $i$ is defined as the fixed point of the function:
\begin{align}
\label{eqn:4}
P_{n-1}(x)&=\frac{E+\frac{\displaystyle\sum_{j\neq i} x_{j(n-1)}+x}{N}}{2}
\end{align}

The solution to this problem is expressed as follows:
\begin{align}
\label{eqn:5}
x_{in}&=\displaystyle\frac{\sum_{j\neq i} x_{j(n-1)}+NE}{2N-1}
\end{align}

In matrix form, the dynamics simplify to:
\begin{align}
\label{eqn:6}
X_n&=G(X_{n-1})=\frac{1}{2N-1} \mathbf{A}X_{n-1}+\frac{N}{2N-1} E
\end{align}

\noindent where $X_n=(x_{1n},\hdots,x_{Nn})$ and $\mathbf{A}$ is the matrix defined by $\mathbf{A}_{ij}=1-\delta_{ij}$. By direct recursion, we have:
\begin{align*}
X_n&=\frac{1}{(2N-1)^{n}}\mathbf{A}^{n}X^{0}+\frac{N}{2N-1}\sum_{m=0}^{n-1} \frac{1}{(2n-1)^m} \mathbf{A}^{m} E
\end{align*}

The power of the matrix $\mathbf{A}$ is given by $\mathbf{A}^n=\displaystyle\frac{1}{N}\left((N-1)^n-(-1)^n\right)\mathbf{J}+(-1)^n\mathbf{I}$, where $\mathbf{J}$ is the matrix of ones and $\mathbf{I}$ is the identity matrix. This implies that
\begin{equation}
\begin{aligned}
\label{eqn:7}
X_n&=\left(\frac{1}{N}\left(\frac{(N-1)^n-(-1)^n}{(2N-1)^{n}}\right)\mathbf{J} +\frac{(-1)^n}{(2N-1)^{n}}\mathbf{I}\right) X^{0}\\
&+\frac{1}{2}\left(\frac{1}{N}\left(\frac{(2N-1)^{n} - 2(N-1)^{n} +(-1)^n}{(2N-1)^{n}}\right)\mathbf{J} +\left(1- \frac{(-1)^n}{(2N-1)^{n}} \right)\mathbf{I}\right) E
\end{aligned}
\end{equation}

The process converges to $E$. The dynamics of the reference price can be deduced as follows:
\begin{align*}
P_{n}&=\frac{\displaystyle\displaystyle\sum_{i=1}^N x_{in}+NE}{2N}\\
&=\frac{(N-1)\sum_{i=1}^N x_{i(n-1)}+(3N^2-N)E}{2N(2N-1)}\\
&=\left(\frac{N-1}{2N-1}\right) P_{n-1}+\left(\frac{N}{2N-1}\right) E
\end{align*}

Using a recursive relationship, we have
\begin{align}
\label{eqn:8}
P_{n}&=\left(\frac{N-1}{2N-1}\right)^n P_{0}+\left(1-\left(\frac{N-1}{2N-1}\right)^n\right) E
\end{align}

\noindent which converges to $E$.\\

When comparing this to the situation in the repeated game under the assumption of common knowledge, we see that after playing their initial strategies $X_0=(x_{10}, \ldots, x_{N0})$ at time $n=0$, the players optimize their choices at time $n = 1$. For each player $i$, the optimal choice is to play $x_{i1}$. However, knowing that every other player $j \neq i$ will play $x_{j1}$, player $i$'s optimal strategy evolves to the next level. Under the common knowledge assumption, and employing reasoning similar to that in the previous section, the players ultimately converge to the limit strategy $X_{\infty} = (E, \ldots, E)$.\\

Now, if we introduce non-negative costs $C_i$ for each player $i$ and let $C=(C_1,\hdots,C_N)$, the player optimal strategy for player $i$ at time $n$ would be given by \mbox{$z_{in}=h_i(x_{in})=\max(C_i,x_{in})$}. In matrix form, where $H=(h_1,\hdots,h_N)$,
\begin{align}
\label{eqn:9}
X_n&=H\circ G(X_{n-1})
\end{align}

Set $X,Y\in[A,B]^N$ and suppose that $H\circ G(X)=G(X)$. Since $H\circ G(Y)\geq G(Y)$ one has
\begin{align*}
|H\circ G(X)-H\circ G(Y)|\leq |G(X)-G(Y)|\leq \frac{\|\mathbf{A}\|}{2N-1}|X-Y|=\left(\frac{N-1}{2N-1}\right)|X-Y|
\end{align*}

Suppose now $H\circ G(X)=C$. In this case, one obtains a null contraction ratio. Therefore, one can deduce that $H\circ G$ is a contraction, which converges to its unique fixed point $x$ defined as follows, given the set $\mathcal{F}=\{j\in \{1,\hdots,N\} \, : \, C_j>E\}$ and $M=\#\mathcal{F}$ ($M$ is the cardinal of $\mathcal{F}$):
\begin{align*}
x&=\displaystyle\begin{cases}
\displaystyle C_i \quad &\text{if} \quad  i\in\mathcal{F}, \\
\displaystyle\frac{\sum_{j\in\mathcal{F}}C_j+NE}{N+M} \quad &\text{otherwise}.
\end{cases}
\end{align*}

This also corresponds to the Nash equilibrium of the reference price guessing game in the presence of non-negative costs.

\vskip 0.5cm

\subsection{The Moroccan public procurement reform\label{The Moroccan public procurement reform}}

\hskip 0.5cm The new reform of the Moroccan Government procurement market \citep{DECRET222431} introduces a revised rule for awarding public contracts. The allocation of contracts is now based on the previously established reference price, which is defined by the formula
\begin{align}
\label{eqn:10}
P&=\frac{E+\displaystyle\frac{\sum_{i=1}^N x_i}{N}}{2}
\end{align}

\noindent where $E$ represents the cost estimate of services established by the contracting authority. The best offer is determined by the following four criteria:

\begin{enumerate}[label=\textit{(\arabic*)}]
\item \label{condition1}The financial offer must fall within the interval \mbox{$[A,B]$}; otherwise, it will be excluded.

\item \label{condition2}The best offer is defined as the one closest to the reference price $P$ by default.

\item \label{condition3}If there are no offers below the reference price, the best offer is the one that is closest to the reference price $P$ by excess.

\item \label{condition4}In the case of a tie, the best offer is determined by drawing lots.\\
\end{enumerate}

This new scenario differs from the previous scenario described in section \ref{The Reference Price Guessing Game}. When rules \ref{condition2}, \ref{condition3}, and \ref{condition4} are applied, the game has no Nash equilibrium. Suppose there exists a Nash equilibrium $X=(x_1,\hdots,x_N)$. If at least one player is not a winner, they have an interest in changing their strategy to win the game, which makes the only possible Nash equilibrium a situation where all players are winners. However, this is not feasible unless $x_1=\hdots=x_N=x$ for some $x$ in $[A,B]$.\\

Suppose $x \geq E$. A player can decrease their price by some $\varepsilon>0$ to be the closest to the reference price by excess and win the bid according to rule \ref{condition3}, thus avoiding the tie situation and rule \ref{condition4}. Conversely, if $x< E$, a player can increase their price by some $\varepsilon>0$ to be the closest to the reference price by default and win the bid according to rule \ref{condition2}, thus avoiding the tie situation and rule \ref{condition4}.

\vskip 0.5cm

\subsubsection{Analysis of Procurement Rules\label{Analysis of Procurement Rules}}

\hskip 0.5cm Our goal in this section is to apply iterated elimination of weakly dominated strategies to the reference price guessing game. Assuming first that the number of bidders is known, we reintroduce the recurrent sequence $(\text{Fix}^n(Z))_{n\in\N}$.\\

Unlike the analysis in section \ref{The Reference Price Guessing Game}, we first note the primacy of rule \ref{condition2} over condition \ref{condition3}. We can derive the following results:

\begin{enumerate}
\item Step 1: Any offer outside the interval $\left[\text{Fix}(A), \text{Fix}(B)\right[$ is weakly dominated. As established in Section \ref{The Reference Price Guessing Game}, the reference price will remain within the interval $\left[\text{Fix}(A), \text{Fix}(B)\right]$. To be closest to the reference price under rule \ref{condition2} or \ref{condition3}, a player's offer should not exceed these bounds. Additionally, the choice $\text{Fix}(B)$ is weakly dominated by $\text{Fix}(B) - \varepsilon$ for some $\varepsilon > 0$, in line with rule \ref{condition2}, if one player at least opts not to play $B$. This condition, however, does not apply to $\text{Fix}(A)$ due to the primacy of rule \ref{condition2}.\\

\item Step 2: Given the results from Step 1 and the revised rational interval $\left[\text{Fix}(A), \text{Fix}(B)\right[$, a similar argument applies, leading to the new rational interval $\left[\text{Fix}^2(A), \text{Fix}^2(B)\right[$.\\

\item $\hdots$\\

\item Step $n$: The rational interval is $\left[\text{Fix}^n(A), \text{Fix}^n(B)\right[$.\\

\item $\hdots$\\

\item At the limit, the rational interval converges to the empty set $\emptyset$.
\end{enumerate} 

If the number of participants is unknown, we can instead use the bounds $\left\{A_n, B_n\right\}$, replacing $\left\{\text{Fix}^n(A), \text{Fix}^n(B)\right\}$, where $A_n=\displaystyle\frac{(2^n-1)E+A}{2^{n}}$ and $B_n=\displaystyle\frac{(2^n-1)E+B}{2^{n}}$. These sequences are the recurrent bounds introduced previously.\\

Remark that $A_n < \text{Fix}^n(A) < \text{Fix}^n(B) < B_n$, and that
\begin{align}
\label{eqn:10bis}
\displaystyle\lim_{n\to \infty} A_n = \lim_{n\to \infty} \text{Fix}^n(A) = \lim_{n\to \infty} \text{Fix}^n(B) = \lim_{n\to \infty} B_n=E.
\end{align}

This implies the following reasoning:

\begin{enumerate}
\item Step 1: Any offer outside the interval $\left]A_1, B_1\right[$ is weakly dominated. By a similar reasoning as previously discussed, offers that fall outside the interval $\left[A_1, B_1\right]$ are weakly dominated. Additionally, because $\text{Fix}(B) < B_1$ and $A_1 < \text{Fix}(A)$, for any number of players $N \geq 2$, the offers $A_1$ and $B_1$ are also weakly dominated.\\

\item Step 2: Based on the findings from Step 1, a similar argument leads to a new rational interval \mbox{$\left]A_2,B_2\right[$}.\\

\item $\hdots$\\

\item Step $n$: Following this iterative process, the rational interval is \mbox{$\left]A_n,B_n\right[$}.\\

\item $\hdots$\\

\item At the limit, considering the limit condition \ref{eqn:10bis}, the rational interval converges to the known $N$ situation, which is the empty set $\emptyset$.
\end{enumerate}

In summary, the preceding analysis confirms that the game has no Nash equilibrium.

\vskip 0.5cm

\begin{remark}{\ }\\
It is important to emphasize that the primacy of rule \ref{condition2} over rule \ref{condition3} encourages bidders to focus on upward analysis (bidding below $E$), unless their costs exceed $E$. Without the common knowledge assumption—considering a situation of bounded rationality—a sort of attractor could form with an illusory equilibrium point at $E$. Refer to the analysis in section \ref{Cognitive hierarchy theory and limited rationality}.
\end{remark}

\vskip 0.5cm

\subsubsection{The best response dynamic under the new rules and the impact of randomness\label{Repeated biding with short memory}}

\hskip 0.5cm Let's consider how players behave in a repeated game under the rules \ref{condition1}--\ref{condition4}. Again, we suppose that each player adjusts their strategy at time $n$ based on the previous trial $n-1$. Player $i$'s optimal strategy $x_{in}$ is the fixed point of the function in equation \ref{eqn:11} minus a random variable $\epsilon_i(\omega)$ (where $\epsilon_i$ represents player $i$'s desire to be closest to the average by default, for some $\epsilon_i > 0$). In matrix form, denoting $X_n(\omega) = (x_{1n}(\omega), \ldots, x_{Nn}(\omega))$, the dynamic is as follows:
\begin{align}
\label{eqn:11}
X_n &= G(X_{n-1}) - \boldsymbol{\epsilon}_n = \frac{1}{2N - 1} \mathbf{A} X_{n-1} + \frac{N}{2N - 1} E - \boldsymbol{\epsilon}_n
\end{align}

\noindent where $\mathbf{A}_{ij} = 1 - \delta_{ij}$ and $\boldsymbol{\epsilon}_n = (\epsilon_{1n}, \ldots, \epsilon_{Nn})$ is the vector of i.i.d. random variables at time $n$. The dynamic is a stationary vector autoregressive model with roots $\displaystyle \left\{-\frac{1}{2N - 1}, \ldots, -\frac{1}{2N - 1}, \frac{N - 1}{2N - 1}\right\}$. By direct recursion, we have:
\begin{align}
\label{eqn:12}
X_n &= \frac{1}{(2N - 1)^{n}} \mathbf{A}^{n} X^{0} + \frac{N}{2N - 1} \sum_{m=0}^{n-1} \frac{1}{(2N - 1)^{m}} \mathbf{A}^{m} E - \sum_{m=0}^{n-1} \frac{1}{(2N - 1)^{m}} \mathbf{A}^{m} \boldsymbol{\epsilon}_{n-m}
\end{align}

We make use of the powers of the matrix $\mathbf{A}$ to write

\begin{equation}
\begin{aligned}
\label{eqn:13}
X_n&=\left(\frac{1}{N}\left(\frac{(N-1)^n-(-1)^n}{(2N-1)^{n}}\right)\mathbf{J} +\frac{(-1)^n}{(2N-1)^{n}}\mathbf{I}\right) X^{0}\\
&+\frac{1}{2}\left(\frac{1}{N}\left(\frac{(2N-1)^{n} - 2(N-1)^{n} +(-1)^n}{(2N-1)^{n}}\right)\mathbf{J} +\left(1- \frac{(-1)^n}{(2N-1)^{n}} \right)\mathbf{I}\right) E\\
&-\sum_{m=0}^{n-1} \frac{1}{(2n-1)^m} \mathbf{A}^{m} \boldsymbol{\epsilon}_{n-m}
\end{aligned}
\end{equation}

Suppose $(\boldsymbol{\epsilon}_n)_{n \geq 1}$ is a sequence of vector i.i.d. normally distributed processes $\mathcal{N}(\boldsymbol{\varepsilon}, \Sigma)$, where $\boldsymbol{\varepsilon} = (\varepsilon_1, \ldots, \varepsilon_N) > 0$ and $\Sigma$ is the diagonal covariance matrix such that $\text{diag}(\Sigma) = (\sigma_1^2, \ldots, \sigma_N^2)$. The process $X_n$ is normally distributed, following the law $\displaystyle\mathcal{N}\left( \left(\mathbf{I}-\frac{1}{2N-1}\mathbf{A}\right)^{-1}\left(\frac{N}{2N-1}E-\boldsymbol{\varepsilon}\right),\Omega\right)$, where \mbox{$\text{vec}(\Omega)=(\mathbf{I}-\mathbf{A}\otimes \mathbf{A})^{-1}\text{vec}(\Sigma)$}, $\text{vec}(\cdot)$ stands for the vectorization of a matrix, and $\otimes$ stands for the Kronecker product.\\

One can deduce the dynamic of the reference price:
\begin{align*}
P_{n}&=\frac{\displaystyle\displaystyle\sum_{i=1}^N x_{in}+NE}{2N}=\frac{E+\frac{\displaystyle\sum_{i=1}^N \frac{\displaystyle\sum_{j\neq i} x_{j(n-1)}+NE}{2N-1} -\epsilon_i}{N}}{2}\\
&=\left(\frac{N-1}{2N-1}\right) P_{n-1}+\left(\frac{N}{2N-1}\right) E - \sum_{i=1}^N\frac{ \epsilon_i}{2N}\\
\end{align*}

$P_n$ is an autoregressive process normally distributed, following the law: \begin{align*}
P_n(\omega)\sim\displaystyle\mathcal{N}\left(E-\frac{2N-1}{2N^2}\sum_{i=1}^N\varepsilon_i,\frac{(2N-1)^2}{N(3N-2)}\sum_{i=1}^N \sigma_i^2\right)
\end{align*}

\vskip 0.5cm

\begin{remark}{\ }\\
\begin{enumerate}
\item The hypothesis of normal distribution of $\boldsymbol{\varepsilon}$ is not necessary.

\item The consideration of non-negative costs produces a sequence of the form $X_n = H\left( G(X_{n-1}) - \boldsymbol{\epsilon}_n\right)$, similar to equation \ref{eqn:9}, and $X_n(\omega)$ is a random variable such that $\mathbb{P}(X_n = a) = 0$ for $C_i > a$.

\item Under the common knowledge hypothesis, the optimal strategy should be the limit process, if there exists
\begin{equation*}
X_{\infty} = E - \lim_{n \to +\infty} \sum_{m=0}^{n-1} \left(\frac{1}{2n-1}\right)^{m} \mathbf{A}^{m} \boldsymbol{\epsilon}_{n-m}.
\end{equation*}

\end{enumerate}
\end{remark}

\vskip 0.5cm

\section{Cognitive Hierarchy Theory and Limited Rationality: Empirical Evidence\label{Cognitive hierarchy theory and limited rationality}}

\hskip 0.5cm In the following section, we develop an empirical model to test the analysis presented in Section \ref{The Keynesian Beauty Contest in Moroccan Public Procurement} for a first play (where players participate in the game for the first time). Since the reform was implemented in 2023, the available data pertains to supply and service contracts that are applying the new rule for the first time. Under rule \ref{condition1}, the bounds are set with $B=+20\%$ above and $A=-25\%$ below the estimated price.\\

We adopt a limited rationality hypothesis, assuming that bidders cannot indefinitely reflect on bids and instead apply the iterated elimination of weakly dominated strategies under common knowledge. Consequently, players are expected to apply this mechanism progressively, starting from step $1$ and transitioning from step $n$ to step $n+1$ according to a geometric process with a parameter $p$ to be determined. As $n$ increases, the probability of a player extending their reasoning to level $n$ diminishes.\\

To test our hypothesis and the conclusions drawn in Section \ref{Analysis of Procurement Rules}, we assume that a player's strategy, without cost constraints for the first bid, is defined by the following stochastic process:
\begin{align}
\label{eqn:14}
x^-_i(\omega) &= \sum_{n=0}^{\infty} \mathbf{1}_{\{G^-(\omega) = n + 1\}} U^-_n(\omega)
\end{align}

\noindent where $G^-$ is a geometric random variable $\mathcal{G}(p^-)$ and $U^-_n$ a uniform variable on $\displaystyle\left[\frac{(2^n-1)E+A}{2^n},E\right]$.

\vskip 0.5cm

\begin{remark}{\ }\\
From discussions with professionals, it appears that bidders in public procurement of services in Morocco are primarily focused on winning the contract rather than on production costs. This trend is expected to continue at least until the VAT (Value Added Tax) withholding at source reform is implemented in July 2024.
\end{remark}

\vskip 0.5cm

$x^-_i$ is a continuous random variable with support in $[A,E]$ (if we suppose that bidders will bid below $E$), its probability density function is given by
\begin{align*}
f^-(y)&=\sum_{n=0}^{\iota(y)}\mathbb{P}\left(G^-=n+1\right)\mathbb{P}\left(U^-_n=y\right)=\sum_{n=0}^{\iota^-(y)}p^-(1-p^-)^{n} \frac{2^n}{E-A}\\
&=\begin{cases}
\displaystyle\frac{(\iota^-(y)+1)}{2(E-A)}  \quad , \quad &\text{if} \quad p^-=\frac{1}{2}\\
\displaystyle\frac{p^-\left(1-\left(2(1-p^-)\right)^{\iota^-(y)+1}\right)}{(2p^- -1)(E-A)} \quad , \quad &\text{otherwise}
\end{cases}
\end{align*}

\noindent where $\iota^-(y)=\displaystyle\arg\max_{n}\left\{n\geq1 \, : \, y\geq\frac{(2^n-1)E+A}{2^n}\right\}$. Recall that \mbox{$A_n=\displaystyle\frac{(2^n-1)E+A}{2^n}$} and \mbox{$B_n=\displaystyle\frac{(2^n-1)E+B}{2^n}$}, for $n\in\N$. One can verify that
\begin{align*}
\int_A^E f^-(y)\,dy &= \sum_{n=0}^{\infty} \int_{A_n}^{A_{n+1}} f^-(y) \,dy\\
&=\begin{cases}
\displaystyle \sum_{n=0}^{\infty} \int_{A_n}^{A_{n+1}} \frac{p^-\left(1-\left(2(1-p^-)\right)^{\iota^-(y)+1}\right)}{(2p^- -1)(E-A)} \,dy=\sum_{n=1}^{\infty} \frac{p^-\left(1-\left(2(1-p^-)\right)^{n}\right)}{(2p^- -1)(E-A)} \frac{E-A}{2^{n}} \quad , &\text{if} \quad p^-\neq \frac{1}{2}\\
\displaystyle \sum_{n=0}^{\infty} \int_{A_n}^{A_{n+1}} \frac{(\iota^-(y)+1)}{2(E-A)} \,dy=\sum_{n=1}^{\infty} \frac{n}{2(E-A)} \frac{E-A}{2^{n}} \quad , &\text{if} \quad p^-= \frac{1}{2}\\
\end{cases}\\
&=1
\end{align*}

One has
\begin{align*}
\mathbb{E}(x^-_i)&=\sum_{n=0}^{\infty}\mathbb{P}\left(G^-=n\right)\mathbb{E}\left(U^-_n\right)=\sum_{n=0}^{\infty}(1-p^-)^{n}p^-\frac{(2^{n+1}-1)E+A}{2^{n+1}}\\
&= \frac{E+p^- A}{1+p^-} \\
\end{align*}

We are able to compute the expected price
\begin{align*}
\mathbb{E}(P)&=\frac{(2+p^-)E+p^- A}{2(1+p^-)} \\
\end{align*}

\noindent and the second moment
{\small
\begin{align*}
\mathbb{E}((x^-_i)^2)&= \int_A^E y^2 f^-(y)\,dy=\sum_{n=0}^{\infty} \int_{A_n}^{A_{n+1}} y^2 f^-(y) \,dy\\
&=\begin{cases}
\displaystyle \sum_{n=0}^{\infty} \int_{A_n}^{A_{n+1}} y^2\frac{p^-\left(1-\left(2(1-p^-)\right)^{\iota^-(y)+1}\right)}{(2p^- -1)(E-A)} \,dy=\sum_{n=1}^{\infty} \frac{p^-\left(1-\left(2(1-p^-)\right)^{n}\right)}{(2p^- -1)(E-A)}\frac{(A_{n})^3-(A_{n-1})^3}{3} \quad , &\text{if} \quad p^-\neq \frac{1}{2}\\
\displaystyle \sum_{n=0}^{\infty} \int_{A_n}^{A_{n+1}} y^2 \frac{(\iota^-(y)+1)}{2(E-A)} \,dy=\sum_{n=1}^{\infty} \frac{n}{2(E-A)}\frac{(A_{n})^3-(A_{n-1})^3}{3} \quad , &\text{if} \quad p^-= \frac{1}{2}\\
\end{cases}\\
&=\frac{4p^-(1+p^-)A^2-2p^-(p^- -5)AE+(9+p^-(p^- -2))E^2}{3(1+p^-)(3+p^-)}
\end{align*}
}

\noindent to get the variance
\begin{align*}
\mathbb{V}(x^-_i)&= \mathbb{E}((x_i^-)^2)-(\mathbb{E}(x_i^-))^2=\frac{p^-(4-p^-(1-p^-))(E-A)^2}{3(3+p^-)(1+p^-)^2}
\end{align*}

\vskip 0.5cm

\begin{figure}[!htb]
\begin{center}
\includegraphics[scale=0.8]{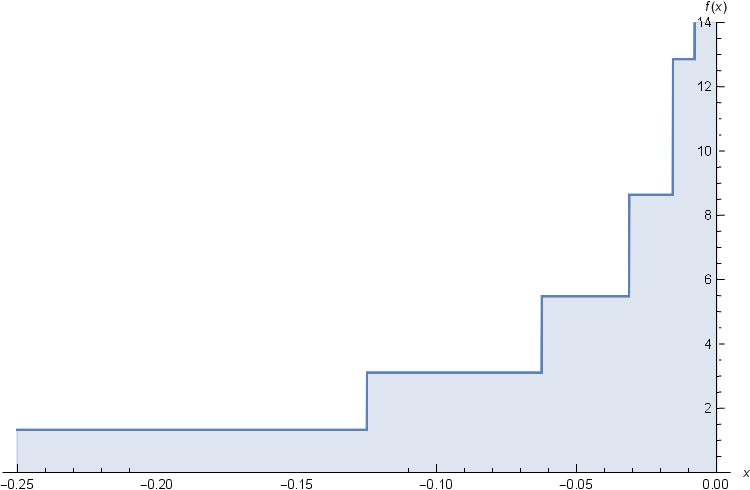}
\caption{The density $f^-(x)$ for $p^-=\displaystyle\frac{1}{3}$.}
\label{Graph1}
\end{center}
\end{figure}

\vskip 0.5cm

One can allow bidding over $E$ using Bernoulli random trial with a probability $q$ that a player will engage in backward reasoning, thus:
\begin{align}
\label{eqn:15}
x_i(\omega)&=\eta(\omega)x^+_i(\omega)+(1-\eta(\omega))x^-_i(\omega)
\end{align}

\noindent where $\eta$ is Bernoulli random variable $\mathcal{B}(q)$, $x^-_i$ is the random process described in the equation \ref{eqn:14}, and $x^+_i$ is a random process similar to $x^-_i$ with support on the upper interval $[E,B]$ such that
\begin{align}
\label{eqn:16}
x^+_i(\omega)&=\sum_{n=0}^{\infty}\mathbf{1}_{\{G^+(\omega)=n+1\}}U^+_n(\omega)
\end{align}

\noindent where $G^+$ is a geometric random variable $\mathcal{G}(p^+)$ and $U^+_n$ is a uniform variable on $\displaystyle\left[E,\frac{(2^n-1)E+B}{2^n}\right]$. The support of the random variable $x^+$ is $[E,B]$, and its probability density function is
\begin{align*}
f^+(y)&=\begin{cases}
\displaystyle\frac{(\iota^+(y)+1)}{2(B-E)}  \quad , \quad &\text{if} \quad p^+=\frac{1}{2}\\
\displaystyle\frac{p^+\left(1-\left(2(1-p^+)\right)^{\iota^+(y)+1}\right)}{(2p^+ -1)(B-E)} \quad , \quad &\text{otherwise}
\end{cases}
\end{align*}

\noindent where $\iota^+(y)=\displaystyle\arg\max_{n}\left\{n\geq1 \, : \, y\leq\frac{(2^n-1)E+B}{2^n}\right\}$, and we can deduce that:
\begin{align*}
\mathbb{E}(x^+_i)&= \frac{E+p^+ B}{1+p^+} \\
\mathbb{V}(x^+_i)&=\frac{p^+(4-p^+(1-p^+))(B-E)^2}{3(3+p^+)(1+p^+)^2}
\end{align*}

Now, let's summarize the properties of the new process $x$ defined in equation \ref{eqn:13}:
\begin{align}
\label{eqn:17}
f(y)&=q^+ f(y)+(1-q) f^-(y)
\end{align}

\begin{align}
\label{eqn:18}
\mathbb{E}(x_i)&=q^+ \frac{E+p^+ B}{1+p^+}+(1-q) \frac{E+p^- A}{1+p^-}
\end{align}

\begin{align}
\label{eqn:19}
\mathbb{V}(x_i)&=\left(q^+\mathbb{E}((x^+_i)^2)+(1-q^+)\mathbb{E}((x^-_i)^2)+q^+(1-q^+)\mathbb{E}(x^+_i)\mathbb{E}(x^-_i)\right) - \left(q^+ \mathbb{E}(x^+_i)+(1-q) \mathbb{E}(x^-_i)\right)^2
\end{align}

We proceed to the estimation of the parameters $(q,p^+,p^-)$ from a set of empirical data $(y_1,\hdots,y_{\mathbb{M}})$. An estimator of $q$ is given by
\begin{align}
\label{eqn:20}
\hat{q}&=\frac{\#\{y_i \, : \, y_i >0\}}{\mathbb{M}}
\end{align}

To estimate the parameter $p^+$ (resp. $p^-$), we split the data set into positive values $(y^+_1,\hdots,y^+_{\mathbb{M}^+})$ and nonpositive values $(y^-_1,\hdots,y^-_{\mathbb{M}^-})$. The first approach is to use the maximum likelihood method:
\begin{align}
\label{eqn:21}
\hat{p}^+&=\arg\max_p\sum_{i=1}^{\mathbb{M}^+}\ln\left(\mathbb{P}(x= y^+_i)\right) \qquad , \qquad \hat{p}^-=\arg\max_p\sum_{i=1}^{\mathbb{M}^-}\ln\left(\mathbb{P}(x= y^-_i)\right)
\end{align}

\noindent which consists of solving a nonlinear problem numerically. The second approach is to apply the moment method
\begin{align}
\label{eqn:22}
\hat{p}^+&=\frac{\bar{y}^+ -E}{B-\bar{y}^+} \qquad , \qquad \hat{p}^-=\frac{E-\bar{y}^-}{\bar{y}^- -A} 
\end{align}

The data consists of $12$ Moroccan public procurement tenders for services, covering the period from October 2023 to April 2024, collected from the electronic government portal \citep{marchespublics}. To normalize the data, we consider the prices' relative deviation from $E$, i.e., $\displaystyle\frac{x_i - E}{E}$. Since the number of bidders is not public information, the adoption of the rational intervals $\displaystyle\left[\frac{(2^n - 1)E + A}{2^n}, \frac{(2^n - 1)E + B}{2^n}\right]$ is convenient.\\

We plot the distribution of the two kinds of bids:

\vskip 0.5cm

\begin{figure}[!htb]
\begin{center}
\includegraphics[scale=0.75]{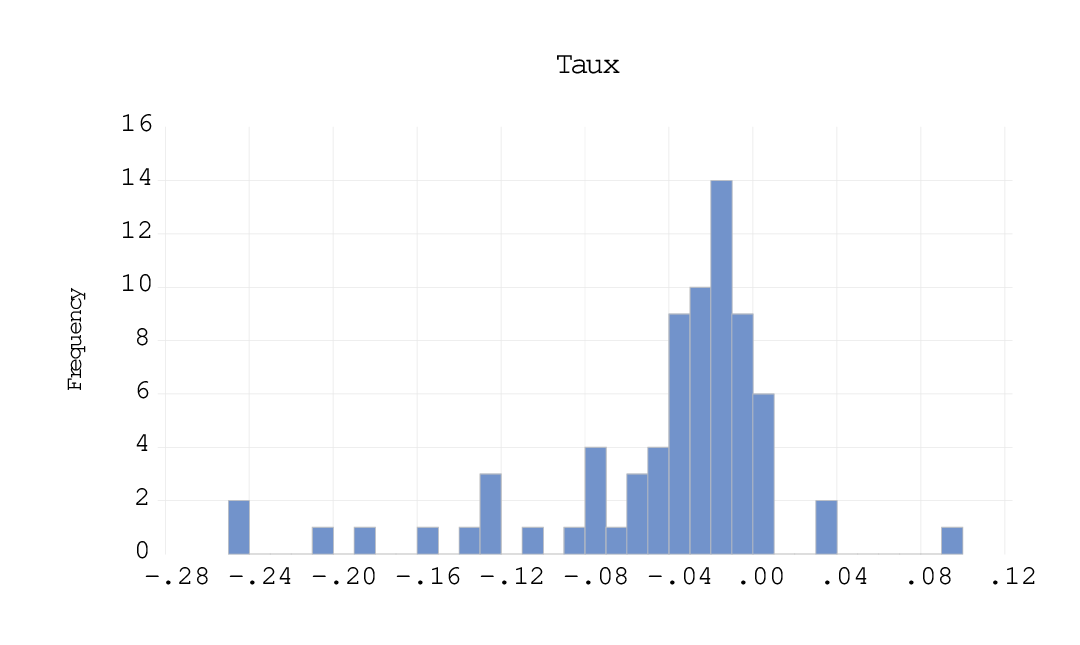}
\caption{Histogram of services bids.}
\label{Graph2}
\end{center}
\end{figure}

\vskip 0.5cm

The table \ref{Table1} summarizes the maximum likelihood estimation results. The estimated transition proportion between the two regimes is $\displaystyle\hat{q} = \frac{8}{74}$:

\vskip 0.5cm

\begin{table}[!htb]
\begin{center}
\begin{tabular}{c| c c c c}
Regime & $\mathbb{M}$ & Log-Likelihood & $\hat{p}$ & $1/\hat{p}$\\
\hline 
\hline
$x^+$ & $8$ & $23.0935$ & $0.236527$ & $4.22785$\\  
\hline
$x^-$ & $66$ & $130.306$ & $0.34617$ & $2.88876$\\   
\end{tabular}
\caption{Maximum likelihood estimation output of the proportion $p$.}
\label{Table1}
\end{center}
\end{table}

\vskip 0.5cm

The estimated value $\hat{q}$ confirms our intuition that the bidders will prioritize bidding under the estimated price $E$, considering rule \ref{condition2}. The estimation emphasizes an asymmetric behavior: $\hat{p}^-$ (resp. $\hat{p}^+$) is about one third ($\frac{1}{3}$) (resp. one quarter ($\frac{1}{4}$)), which means that at each level of reasoning, two out of three (resp. three out of four) individuals decide to proceed to the next level. The average reasoning level is about $3$ (resp. $4$).
\\

Next, we use the moment method estimator to establish a bilateral confidence interval for both $p^+$ and $p^-$ at the level $1 - \alpha = 0.99$ in each regime. Note that $\mathbb{E}(x_i^-)$ (resp. $\mathbb{E}(x_i^+)$) is an increasing (resp. decreasing) function of $p$:

\vskip 0.5cm

\begin{table}[!htb]
\begin{center}
\begin{tabular}{c| c c c c c}
Regime & $\mathbb{M}$ & $\hat{p}_{\frac{\alpha}{2}}$ & $\hat{p}$ & $\hat{p}_{1-\frac{\alpha}{2}}$\\
\hline 
\hline
$x^+$ & $8$ & $0.168246$ & $0.134076$ & $0.109883$\\  
\hline
$x^-$ & $66$ & $0.195254$ & $0.24788$ & $0.322176$\\
\end{tabular}
\caption{Confidence interval estimation at level $99\%$ for $p$.}
\label{Table2}
\end{center}
\end{table}

\vskip 0.5cm

\section{Moroccan public markets principles and manipulations\label{Moroccan public markets principles and manipulations}}

\hskip 0.5cm An advantage of the first-price sealed-bid auction lies in its resistance to manipulation. Each bidder submits a sealed bid without knowledge of the others, and the lowest bidder wins, receiving their bid amount. However, when analyzing the rules introduced by the new reform, one could suspect the possibility of manipulation, especially considering the sensitivity of the mean to extreme data.\\

Consider a coalition $\mathcal{C}$ consisting of $l$ out of the $N$ bidders (if $N$ is unknown, it can be estimated or inferred from past plays). Given our results in section \ref{Cognitive hierarchy theory and limited rationality}, the player can construct an estimation of the vector $\boldsymbol{p}=(q,p^+,p^-)$. Denote by $\boldsymbol{\hat{p}}=(\hat{q},\hat{p}^+,\hat{p}^-)$ this estimation.\\

The winning strategy is as follows: the first $l-1$ coalition members choose the upper bound $B$, while the other coalition members choose $x$, the solution of
\begin{align}
\label{eqn:23}
x&= \frac{E+\displaystyle\frac{(N-l)\mathbb{E}\left(x_i(\boldsymbol{\hat{p}})\right)+(l-1)B+x}{N}}{2}
\end{align}

that is
\begin{align}
\label{eqn:24}
x^{\star}&=\frac{(N-l)\mathbb{E}\left(x_i(\boldsymbol{\hat{p}})\right)+(l-1)B+NE}{2N-1}
\end{align}

There is a risk of the contract being stolen by a downward player represented by the probability $\mathbb{P}\left(x_j(\boldsymbol{p})\in\left]x^{\star},P\right] \, : \, j\not\in \mathcal{C}\right)$.
\\

For example, using the data of section \ref{Cognitive hierarchy theory and limited rationality}, and letting $N=10$, one can calculate the winning strategy $x^{\star}$ and an upper bound estimate of the probability of a stolen win: $\mathbb{P}\left(x_j(\boldsymbol{\hat{p}})\in\left]x^{\star},\frac{B+E}{2}\right] \, : \, j\not\in\mathcal{C}\right)$.

\vskip 0.5cm

\begin{table}[!htb]
\begin{center}
\begin{tabular}{c| c c c c c}
$l$ & $2$ & $3$ & $4$ & $5$ & $6$\\
\hline 
\hline
$\displaystyle x^{\star}$ & $-0.01188$ & $0.00145$ & $0.01478$ & $0.02811$ & $0.04143$\\
\hline
$\displaystyle\mathbb{P}\left(x_j(\boldsymbol{\hat{p}})\in\left]x^{\star},\frac{B+E}{2}\right] \right)$ & $0.25734$ & $0.07229$ & $0.03634$ & $0.02268$ & $0.01871$\\
\hline
$\displaystyle\mathbb{P}\left(x_j(\boldsymbol{\hat{p}})\in\left]x^{\star},\frac{B+E}{2}\right] \, : \, j\not\in\mathcal{C}\right)$ & $0.94892$ & $0.52782$ & $0.30935$ & $0.20500$ & $0.17209$\\
\end{tabular}
\caption{$x^{\star}$ as a function of the coalition cardinal.}
\label{Table3}
\end{center}
\end{table}

\vskip 0.5cm

One can observe that winning the game is a possibility, and the estimated chance of winning increases as $l$ grows. While the outcome is not certain, this possibility could encourage the formation of coalitions.

\vskip 0.5cm

\subsection*{The median solution}

\hskip 0.5cm The median is a natural candidate to neutralize the mean's sensitivity to extreme data. One can suggest replacing the mean in the reference price formula with the median in the following manner:
\begin{align}
\label{eqn:25}
P&=\frac{E+\displaystyle q_{\frac{1}{2}}(x_1,\hdots,x_N)}{2}
\end{align}

\noindent where $\displaystyle q_{\frac{1}{2}}$ stands for the median of the offers $x_1,\hdots, x_N$.\\

In this scenario, $E$ regains its role as the unique Nash equilibrium under the rules \ref{condition1}, \ref{condition2}, \ref{condition3}, and \ref{condition4}, since no one can individually influence the median. Any situation other than a tie is also not a Nash equilibrium, and every tie different from $E$ is not a Nash equilibrium either, as every player can secure the contract by changing their price.\\

An iterative elimination of weakly dominated strategies can be performed in the same manner as described in section \ref{The Keynesian Beauty Contest in Moroccan Public Procurement}. However, the game becomes almost surely manipulable under a stringent condition: the formation of a coalition comprising more than half the players.
\begin{enumerate}
\item If $N$ is odd, with a coalition of $\displaystyle \frac{N+3}{2}$ players, $\displaystyle\frac{N+1}{2}$ players choose the maximum value $B$, while the $\displaystyle\frac{N+3}{2}$th player wins the contract by playing the reference price $P=\displaystyle\frac{B+E}{2}$.

\item If $N$ is even, with a coalition of $\displaystyle \frac{N+4}{2}$ players, $\displaystyle\frac{N+2}{2}$ players choose the maximum value $B$, and the $\frac{N+2}{2}$th player wins the game by playing the reference price $P=\displaystyle\frac{B+E}{2}$.
\end{enumerate}

\vskip 0.5cm

\section{Conclusion}

\hskip 0.5cm This paper challenges the common knowledge assumption within auction theory by examining the "reference price guessing game". This game, generously introduced by the General Treasury of the Kingdom of Morocco, engages real bidders whose decisions significantly impact their businesses, all without any Hawthorne effect\footnote{The Hawthorne effect is a type of human behavior reactivity in which individuals modify an aspect of their behavior in response to their awareness of being observed\citep{Fox2008}.}.\\

Our findings offer new insights into sealed auction dynamics within Moroccan public procurement. We provided evidence of limited rationality by estimating the average level of reasoning among bidders. Our analysis shows that players generally do not extend their reasoning beyond levels three or four, making the common knowledge hypothesis unrealistic. Furthermore, we explored the implications of coalition formation due to the mean's sensitivity to extreme data, highlighting the risk of manipulation that could arise in such a framework. We propose using the median as an alternative measure to mitigate this sensitivity, contributing to auction theory and public procurement literature.\\

Our approach faces some limitations. One notable aspect we could not empirically test is the potential for learning within the game and the possibility of an autoregressive dynamic in the bids. The current dataset, which is limited to just 12 Moroccan public procurement tenders, restricts our ability to draw broader conclusions about bidding behavior over time. Furthermore, the analysis is based on a specific regulatory context, which may limit the generalizability of our findings to other environments.\\

Future research could focus on the dynamics of learning in repeated bids, potentially employing longitudinal data to explore how bidders adapt their strategies over time. Investigating the formation of coalitions further could also yield valuable insights, especially in different regulatory contexts or industries. Additionally, assessing the long-term effects of Morocco's new public procurement market reform on efficiency and bidder behavior could provide a deeper understanding of its implications for public services and market competitiveness.\\

The rationale behind Morocco's new public procurement market reform is worth questioning. Unlike the old rule, where the bidder with the lowest price wins, the new rule selects a "random" (in efficiency sense) winner with the "best price." While this could potentially ensure higher profits and prevent monopolization by a single low-cost bidder, it raises concerns about the efficiency of public services and whether the reform genuinely promotes efficient companies. Moreover, it may encourage manipulation and coalition formation.

\clearpage

\section*{Declarations}

\subsection*{Conflict of interest}

The author declares that they have no conflict of interest.

\subsection*{Authors' contributions}

The author carried out the study conception and design. Material preparation, data collection and analysis were performed by the author. The manuscript was written by the author.

\subsection*{Funding}

No funding was received for conducting this study.

\subsection*{Availability of data and materials}

The data used comes from Moroccan Public Procurement Portal freely accessible at :\\
\mbox{\url{https://www.marchespublics.gov.ma}}.

\subsection*{Declaration of generative AI and AI-assisted technologies in the writing process}

During the preparation of this work the author used ChatGPT in order to improve the text. After using this tool/service, the author reviewed and edited the content as needed and takes full responsibility for the content of the publication.

\bibliographystyle{apalike}
\bibliography{BibliographieEco}

\end{document}